\documentclass[onecolumn,floatfix,prb,aps,showpacs]{revtex4}
\usepackage{graphicx,amsmath,amssymb,color}
\usepackage{nicefrac}
\usepackage{url}
\usepackage[titletoc,title]{appendix}

\newcommand{\be}{\begin{equation}}
\newcommand{\ee}{\end{equation}}

\newcommand{\ba}{\begin{eqnarray}}
\newcommand{\ea}{\end{eqnarray}}

\begin{document}

\title{Fractionally charged skyrmions in fractional quantum Hall effect}
\author{Ajit C. Balram,$^1$ U. Wurstbauer,$^2$ A. W\'ojs,$^3$ A. Pinczuk$^4$ \& J. K. Jain$^1$}
\affiliation{$^1$Department of Physics, 104 Davey Lab, Pennsylvania State University, University Park, Pennsylvania 16802, USA}
\affiliation{$^2$Walter Schottky Institut and Physik-Department, Am Coulombwall 4a, Technische Universit\"at M\"unchen, D-85748 Garching, Germany and and Nanosystems Initiative Munich (NIM), Munich 80799, Germany.}
\affiliation{$^3$Department of Theoretical Physics, Wroc{\l}aw University of Technology, 50-370 Wroc{\l}aw, Poland}
\affiliation{$^4$Department of Applied Physics and Applied Mathematics and Department of Physics, Columbia University, New York 10027, USA}

\begin{abstract}
The fractional quantum Hall effect has inspired searches for exotic emergent topological particles, such as fractionally charged excitations, composite fermions, abelian and
nonabelian anyons and Majorana fermions. Fractionally charged skyrmions, which support both topological charge and topological vortex-like spin structure, have also been predicted to occur in the vicinity of 1/3 filling of the lowest Landau level. The fractional skyrmions, however, are anticipated to be exceedingly fragile, suppressed by very small Zeeman energies. Here we show that, slightly away from 1/3 filling, the smallest manifestations of the fractional skyrmion exist in the excitation spectrum for a broad range of Zeeman energies, and appear in resonant inelastic light scattering experiments as well-defined resonances slightly below the long wavelength spin wave mode. The spectroscopy of these exotic bound states serves as a sensitive tool for investigating the residual interaction between composite fermions, responsible for delicate new fractional quantum Hall states in this filling factor region.
\end{abstract}


\date{\today}
\maketitle

Skyrmions represent vortex-like spin structures in two dimensions, which are two-dimensional stereographic projections of the spin hedgehog on a sphere. In a pioneering work, Sondhi {\em et al.}\cite{Sondhi93} predicted integrally charged skyrmions for the quantum Hall ferromagnet near filling factor $\nu=1$. These arise because the $\nu=1$ integer quantum Hall effect (IQHE) state exhibits spontaneous ferromagnetism even in the absence of a Zeeman energy with the remarkable property that the addition or removal of a single electron causes a macroscopic number of spin-flips\cite{Sondhi93}. For non-zero Zeeman energies, the number of spin flips depends on the competition between the exchange and the Zeeman energies\cite{Fertig94}, i.e. on the parameter $\kappa=E_{\rm Z}/(e^2/\epsilon \ell)$, which characterizes the strength of the Zeeman splitting $E_{\rm Z}=g\mu_{\rm B}B$. (Here $\epsilon$ is the dielectric constant of the background material, $B$ is the perpendicular magnetic field, $\ell=\sqrt{\hbar c/eB}$ is the magnetic length, $g$ is the Land\'e g-factor, and $\mu_{\rm B}$ is the Bohr magneton. For the parameters of GaAs with the magnetic field specified in units of Tesla (T), we have $\kappa\approx 0.006 \sqrt{B}$.)  The skyrmion physics is relevant for $\kappa\lesssim 0.05$ ($B\lesssim 70$T for GaAs), and has been confirmed experimentally\cite{Barrett95,Schmeller95}.

The fractional quantum Hall effect \cite{Tsui82} (FQHE) arises due to the formation of composite fermions (CFs), which are topological bound states of electrons and an even number ($2p$) of quantized vortices \cite{Jain89}. Composite fermions experience an effective magnetic field and form Landau-like levels called $\Lambda$ levels ($\Lambda$Ls). Their filling factor $\nu^*$ is related to the electron filling factor $\nu$ by the expression $\nu=\nu^*/(2p\nu^*\pm 1)$. The FQHE states at $\nu=n/(2n\pm 1)$ are manifestations of IQHE of composite fermions and one may expect fractionally charged skyrmions close to composite fermion filling $\nu^*=1$, which corresponds to the $\nu=1/3$ state\cite{Laughlin83}. Such fractional skyrmions (FSs) were already predicted in the work of Sondhi {\em et al.}\cite{Sondhi93}, and their existence was subsequently verified in detailed microscopic calculations\cite{Kamilla96,Wojs02,Doretto05}. These calculations indicated, however, that the FSs are much more delicate than the integral skyrmions near $\nu=1$, because the exchange interaction between composite fermions is much weaker than that between electrons. It was estimated that FSs should occur only below $\kappa\approx 0.009$ ($B\lesssim 2.5$T for GaAs). Consequently, for typical experimental parameters, when the filling factor is varied away from $\nu^*=1$ (or from $\nu=1/3$), trivial quasiparticles, namely isolated CF particles or holes, are produced rather than FSs. The FSs have been probed experimentally by suppressing $g$ through application of hydrostatic pressure\cite{Leadley97}. Certain $E_{\rm Z}$ dependencies of the excitations\cite{Detlefsen06,Groshaus08} at $\nu=1/3$ have also been interpreted in terms of skyrmion physics, but it is unclear how skyrmions may occur at the high $\kappa$ values of these experiments, and an alternative explanation of the observations has been proposed\cite{Murthy09}. The binding energy of FSs has not been measured so far, which would be important for a convincing observation.

This work is concerned with minimal FSs, namely the skyrmions for which a CF particle or hole is dressed by a single additional spin-flip exciton (SFE). We show theoretically that such skyrmions exist in the excitation spectrum just below the Zeeman energy for a broad range of $\kappa$ at filling factors slightly away from $\nu=1/3$. These are accessible in resonant inelastic light scattering (RILS), because a photo-excited SFE that can bind, for $\nu>1/3$ ($\nu<1/3$), with a pre-existing CF particle (hole) to produce a negatively (positively) charged FS$^{-}$ (FS$^{+}$). We identify certain modes observed in RILS experiments\cite{Gallais06,Dujovne03} with the minimal FSs, supporting this identification by a detailed analysis of the experimental data, which shows qualitative and quantitative agreement between theory and experiment. In particular, the measured binding energies of the FS$^{\pm}$ are seen to be in excellent agreement with the calculated binding energies.

\section{Results}

{\bf Theory:} Employing a combination of exact and CF diagonalization methods, we evaluate the binding energy of the minimal FS, i.e. the amount by which it lies below the Zeeman energy, and estimate corrections due to finite quantum well thickness. We consider filling factors close to $\nu=1/3$, where the density of CF particles or holes is dilute and it suffices to consider a single CF particle or hole. The CF hole resides in the spin-up lowest $\Lambda$L (0$\uparrow$ $\Lambda$L) for all $\kappa$ (Fig.~\ref{Fig1}(a)), whereas the CF particle can reside either in spin-down lowest $\Lambda$L (0$\downarrow$ $\Lambda$L) for small $\kappa$ (Fig.~\ref{Fig1}(c)) producing a partially polarized state, or in spin-up second $\Lambda$L (1$\uparrow$ $\Lambda$L) for large $\kappa$ (Fig.~\ref{Fig1}(e)) producing a fully spin polarized state.

We use the spherical geometry\cite{Haldane83}, in which $N$ electrons move on the surface of a sphere, exposed to a radial magnetic field that produces a flux of $2Q \phi_0$ through the surface of the sphere, where $2Q$ is a positive integer and $\phi_0=hc/e$ is a unit flux quantum. The distance between the electrons is defined as the chord distance on the sphere -- whether the chord or arc distance is chosen is unimportant because we evaluate the thermodynamic limit of the energy. The $\nu=1/3$ state occurs at $2Q=3N-3$ and has the spin quantum number $S=N/2$. A single CF hole occurs at $2Q=3N-2$ with the spin of the bare CF particle (Fig.~\ref{Fig1}(a)) also given by $S=N/2$. The CF particle at $2Q=3N-4$ can go either into 0$\downarrow$ $\Lambda$L (Fig.~\ref{Fig1}(c)), with spin $S=N/2-1$, or into 1$\uparrow$ $\Lambda$L (Fig.~\ref{Fig1}(e)), with spin $S=N/2$.  The red dashes in Fig.~\ref{Fig1} (g), (h) and (i) are obtained by exact diagonalization in these spin sectors. We can also construct explicit wave functions for these states in the CF theory, which for the CF hole has the form $\Psi_{1/3}^{\rm CF-hole}={\cal P}_{\rm LLL}\prod_{j<k}(u_jv_k-v_ju_k)^2 \Phi_1^{\rm hole}$ where $\Phi_1^{\rm hole}$ is the known wave function of a single hole at $\nu=1$, $u=\cos(\theta/2)\exp(i\phi/2)$ and $v=\sin(\theta/2)\exp(-i\phi/2)$ are spinor coordinates, $\theta$ and $\phi$ are the polar and azimuthal angles on the sphere, and ${\cal P}_{\rm LLL}$ represents projection into the lowest Landau level (LLL). Wave functions for spin-conserving and spin-reversed CF particles are constructed analogously. The red dots in panels (g), (h) and (i) of Fig.~\ref{Fig1} are the Coulomb energies of these wave functions. 

To consider the FSs, we next consider states containing an additional SFE, shown in Fig.~\ref{Fig1} (b), (d) and (f) and consider the sector $\Delta S=-1$, where $\Delta S$ is measured relative to the ground state. We show the spectra in these spin sectors in Fig.~\ref{Fig1} in panels (g), (h) and (i). The black dashes show the exact Coulomb spectra obtained by numerical diagonalization. The black dots are the spectra obtained by the method of CF diagonalization\cite{Mandal02}. For the latter, we first construct a basis of all states in the relevant spin sector at $2Q^*=2Q-2(N-1)$ (which is the effective flux experienced by composite fermions), denoted by $\{\Phi^{\alpha}_{Q^*}\}$, where $\alpha$ labels different basis functions. We then composite fermionize this basis to obtain the correlated CF basis at $2Q$, given by $\{\Psi_{Q}^\alpha={\cal P}_{\rm LLL}\prod_{j<k}(u_jv_k-u_kv_j)^2\Phi^{\alpha}_{Q^*}\}$. We finally diagonalize the Coulomb interaction in this basis to obtain eigenenergies and eigenfunctions.

In both the exact and the CF spectra, we find that a FS bound state (highlighted in yellow in panels Fig.~\ref{Fig1}(g) and (h)) is produced when a CF particle in the $0$$\downarrow$ $\Lambda$L or a CF hole in the $0$$\uparrow$ $\Lambda$L is dressed by the spin flip exciton. No such bound state is produced for the fully spin polarized state at $\nu>1/3$, i.e. for a CF particle in the $1$$\uparrow$ $\Lambda$L. The exact density profiles of the FSs are seen to be qualitatively different, and much smoother, than those of the CF particle or CF hole (see Figs.~\ref{Fig2} and \ref{Fig3}), which is what results in the lowering of the Coulomb energy. Despite the remarkably different structures, they all carry a precise fractional charge of magnitude $e/3$. Fig.~\ref{Fig4} shows the thermodynamic extrapolation of the binding energy of the fractional skyrmions, denoted $E^{\pm}_{\rm b}$, obtained from exact diagonalization results for finite systems. (The energy of FS$^{\pm}$ is given by $E_{\rm Z}-E^{\pm}_{\rm b}$.) The thermodynamic limits for the binding energies are determined to be $E^+_{\rm b}=0.0096(2) e^2/\epsilon \ell$ and $E^-_{\rm b}=0.0052(2) e^2/\epsilon \ell$ for a system with zero thickness and no Landau level (LL) mixing.

The interpretation of the FS$^\pm$ as bound states of three composite fermions (see panels (b) and (d) of Fig.~\ref{Fig1}) is confirmed by: the close agreement between the energies of the exact and the CF wave functions (i.e. the dashes and the dots in Fig.~\ref{Fig1}); by a comparison of the density profiles of the exact and CF wave functions shown in Fig.~\ref{Fig5}; and the high overlap of $\sim$0.99 between the exact and the CF wave functions for $N=12$.

For an accurate quantitative comparison with experiment, we have estimated corrections due to finite transverse width of the quantum well wave function. We first use a local density approximation (LDA) \cite{Ortalano97} to obtain the transverse wave function $\xi(z)$. The effective two-dimensional interaction is given by:
$V^{\text{eff}}({\bf r}) =\frac{e^2}{\epsilon}\int dz_{1}\int dz_{2} \frac{|\xi(z_{1})|^2|\xi(z_{2})|^2}{\sqrt{r^2+(z_{1}-z_{2})^2}}$,
where $z_{1}$ and $z_{2}$ are the coordinates perpendicular to the plane containing the electrons. At short distances this interaction is softer than the Coulomb interaction. For the FSs, the change in the energy due to finite width is shown in the inset of Fig.~\ref{Fig4}. We use the CF theory for obtaining the corrections due to finite width because it is possible to go to larger systems in CFD than in exact diagonalization; the use of the CF theory is justified given the above result showing the accuracy of the CF theory. The finite size variations preclude a clean extrapolation to the thermodynamic limit $1/N\rightarrow 0$, but it is clear that the binding energies for the FSs are reduced only by a small amount. We take the average of all points in the inset of Fig.~\ref{Fig4} as a measure of the reduction in the FS binding energy due to finite width, which gives for FS$^{+}$ and FS$^{-}$ energy reductions of 0.0013 (0.0005) and 0.0010 (0.0001) $e^2/\epsilon \ell$, respectively (with the error given by the standard deviation). We apply this correction to the binding energies obtained from exact diagonalization.

{\bf Experiment:} The experimental results presented in this work are from resonant inelastic light scattering (RILS) on a high quality GaAs single quantum well (SQW) of width 33 nm, electron density n=5.5$\times$10$^{10}$~cm$^{-2}$ and low temperature mobility $\mu$=7.2$\times$10$^6$~cm$^2$/Vs. The magnetic field perpendicular to the sample is B=B$_{\rm Total}$cos$\theta$, where $\theta$ is the tilt of the sample with respect to the direction of the total magnetic field $B_{\rm Total}$. The filling factor and magnetic length depend on the perpendicular field $B$ whereas the Zeeman energy on the total field B$_{\rm Total}$, and thus tilting can be used to vary the parameter $\kappa$ ($\kappa$ is defined as the ratio of the Zeeman to Coulomb energy). Measurements were taken at two tilt angles, $\theta$=30$^{o}\pm$2$^{o}$ and $\theta$=50$^{o}\pm$2$^{o}$, which correspond to $\kappa\approx 0.018$ and $\kappa\approx 0.023$. RILS spectra spectra were obtained by tuning the incident laser photon energy $E_{\rm laser}$ to be close to the fundamental optical gap of GaAs to enhance the light scattering cross-section. To identify all modes it is important to scan over a range of energies of the incoming laser photon, because modes are picked out by resonant Raman scattering most prominently in a narrow range of parameters where the resonance condition is best satisfied. As seen in Figs.~\ref{Fig6}(a) and \ref{Fig6}(b), three modes can be identified for 30$^\circ$ tilt, whereas only two are seen for 50$^\circ$ tilt. In many cases the number of modes and their energies are evident without fitting (see for example Fig. 2 in Ref. \cite{Gallais06} and Fig. 3 in Ref. \cite{Dujovne03}). In general, a detailed line-shape analysis is necessary to determine the number and the energies of the observed modes. To this end, we determine the least number of Lorentzians that provide a reasonably good fit to the observed Raman line shapes, with the centers of Lorentzians giving the energies of the modes. Some representative fits  for 30$^{o}$ tilt are shown in Fig.~\ref{Fig7}, where at least three Lorentzians are needed for a good fit to the observed line shape. For 50$^\circ$ tilt, a fit with two Lorentzians is found to be satisfactory.

Figs.~\ref{Fig1}(g)-(l) show energies of experimental modes observed in RILS at two tilt angles between magnetic field direction and the plane-normal. For 50$^\circ$ til the results are taken from Gallais et al.\cite{Gallais06}. For 30$^\circ$ tilt, the experimental points shown in Fig.~\ref{Fig1}(k) are deduced from the RILS spectra of Dujovne {\em et. al.} \cite{Dujovne03}, but the detailed line shape analysis performed here gives more accurate energies than those quoted in that work. 

\section{Discussion}

\begin{figure}
\begin{center} 
\includegraphics[width=7in, height=3.5in]{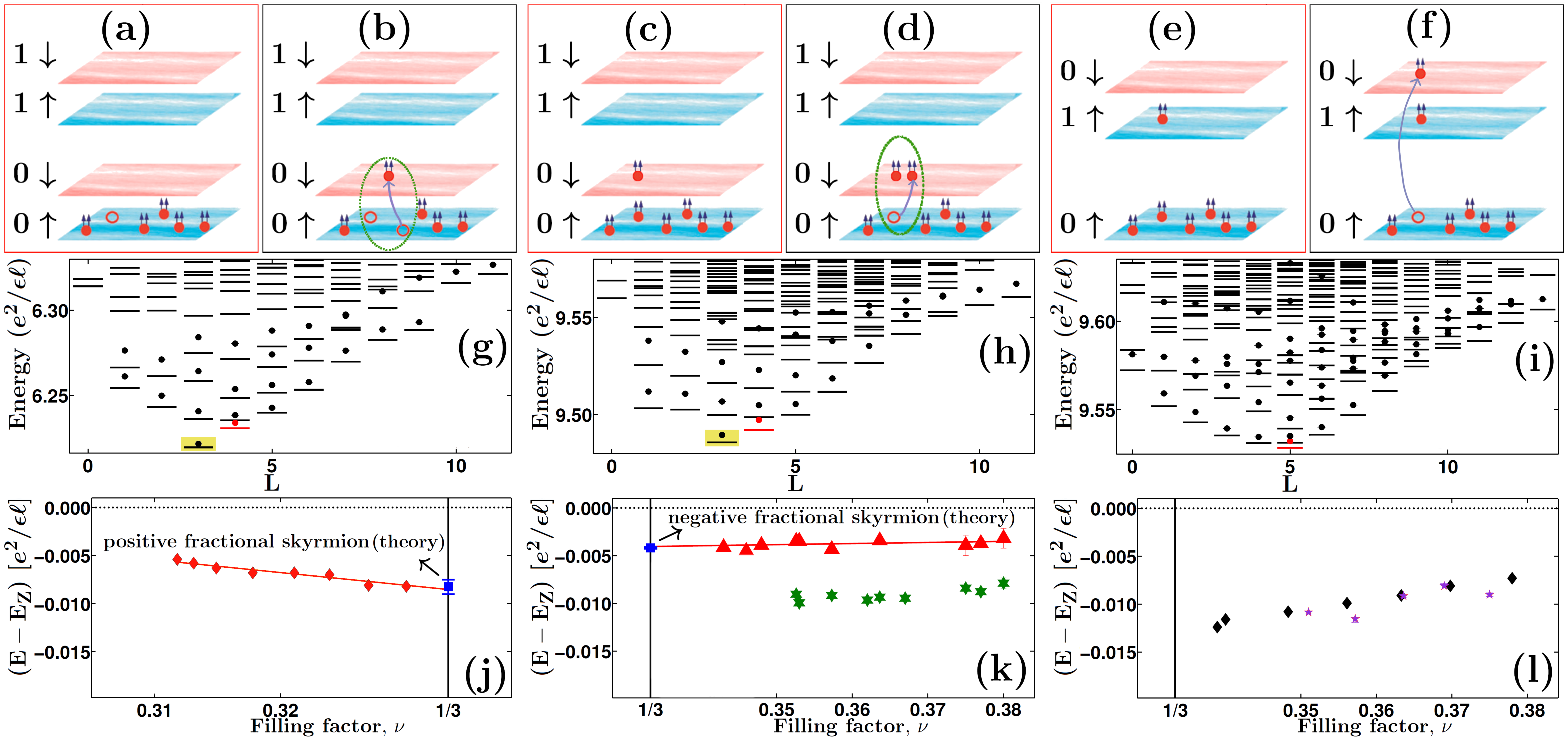}
\end{center}  
\caption{\label{Fig1}{\bf Comparison between theory and experiment.} (a) shows the ground state for the fully polarized state for $\nu\lesssim 1/3$ with a single CF hole (empty red circle) in the spin-up lowest $\Lambda$L ($0$$\uparrow$); (b) shows an additional spin-flip exciton (SFE) that binds with the hole to produce a minimal positively charged fractional skyrmion (FS). (c) shows the state for $\nu\gtrsim 1/3$ with a single CF particle in the spin-down lowest $\Lambda$L ($0$$\downarrow$), and (d) has an additional SFE. (e) has a CF particle in the spin-up second $\Lambda$L ($1$$\uparrow$), and (f) has an additional SFE. The composite fermions are shown as particles with two arrows, representing bound vortices, and their up and down spin $\Lambda$Ls are shown as shaded blue and red rectangles, respectively. In (g), (h) and (i), the red dashes (dots) show the exact (CF) energies of the ground states containing a single CF particle or hole (as shown in (a), (c) and (e)), and the black symbols show the spectrum obtained when an additional SFE is created (as shown (b), (d), and (f)). The spherical geometry is used for calculations; the (g) is for 8 particles subjected to 22 flux quanta (a flux quantum is defined as $\phi_0=hc/e$), and (h) and (i) correspond to 10 particles in 26 flux quanta. The panels (j), (k) and (l) show the experimentally measured energies of modes below the Zeeman energy. The theoretical energy of the FSs in the dilute limit of $\nu\rightarrow 1/3$ including finite width correction is also shown by blue square. The panels (j) and (l) are for 50$^\circ$ tilt, whereas (k) is for 30$^\circ$ tilt. All energies in (j), (k) and (l) are shown relative to the Zeeman energy, in units of $e^2/\epsilon \ell$, where $\epsilon$ is the dielectric constant of the material and $\ell$ is the magnetic length. The modes depicted by red symbols are assigned to fractional skyrmions, green stars in panel (k) to the excitation shown in Fig. 8(d), and the black diamonds and purple stars in panel (l) to the excitation shown in Fig. 8(i).. The theoretical error bars arise from the uncertainty in the Monte Carlo calculations and thermodynamic extrapolations, and the experimental error bar reflects the uncertainty in the Lorentzian fits.}
\end{figure}

\begin{figure}
\centering
\includegraphics[width=3.5in]{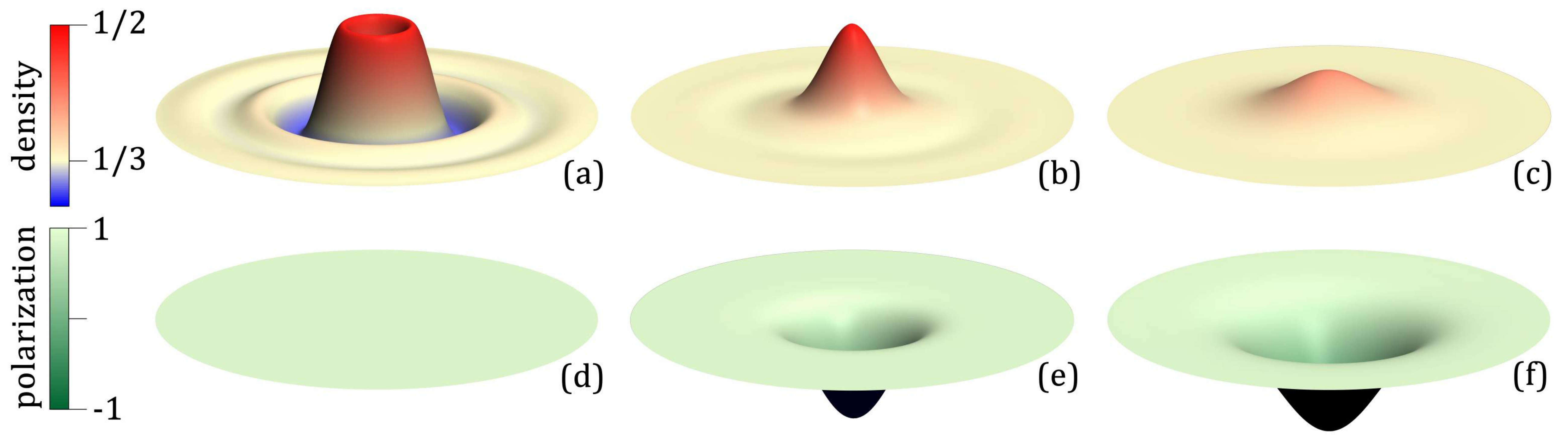}
\caption{\label{Fig2}{\bf Contrasting negatively charged skyrmion with CF particle.} Panels (a), (b) and (c) show charge density profiles of a spin-conserving CF particle, a spin-reversed CF particle, and a negatively charged fractional skyrmion. Their spin polarization, defined by $(\rho_\uparrow(r)-\rho_\downarrow(r))/(\rho_\uparrow(r)+\rho_\downarrow(r))$ where $\rho_{\uparrow}(r)$ and $\rho_{\downarrow}(r)$ are the spatial densities of spin-up and spin-down composite fermions, is shown in panels (d), (e) and (f) respectively. The minimum/maximum values of the color bars in each panel are: (a) 0.303/0.453, (b) 0.333/0.456, (c) 0.333/0.391, (d) 1.000/1.000, (e) $-$0.352/1.000, (f) $-$0.512/1.000. The disk has a radius of 12.5 $\ell$.}
\end{figure}

\begin{figure}
\centering
\includegraphics[width=3.5in]{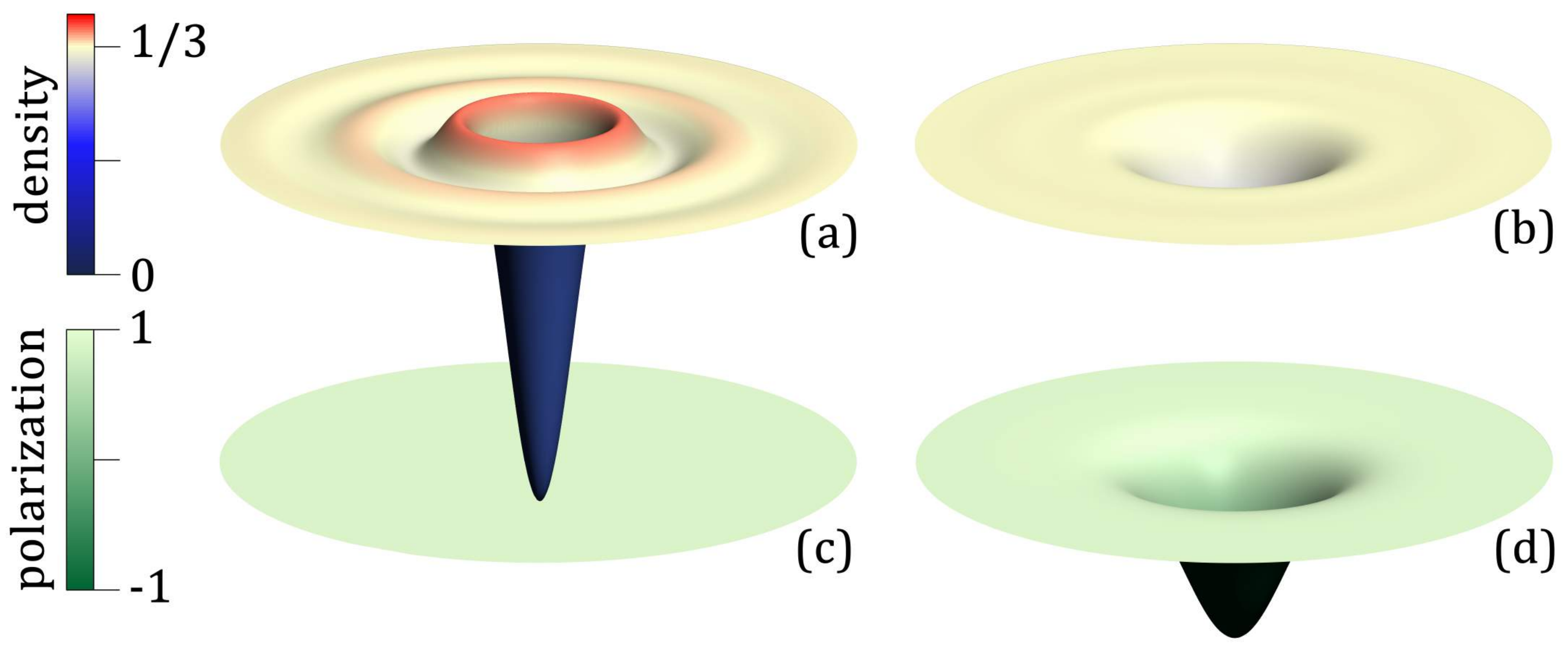}
\caption{\label{Fig3}{\bf Contrasting the positively charged skyrmion with the CF hole.} Panels (a) and (b) show charge density profiles of
a CF hole and a positively charged fractional skyrmion. Their spin polarization, defined by $(\rho_\uparrow(r)-\rho_\downarrow(r))/(\rho_\uparrow(r)+\rho_\downarrow(r))$ where $\rho_{\uparrow}(r)$ and $\rho_{\downarrow}(r)$ are the spatial densities of spin-up and spin-down composite fermions, is shown in panels (c) and (d) respectively. The minimum/maximum values of the color bars in each panel are: (a) 0.006/0.357, (b) 0.266/0.333, (c) 1.000/1.000, (d) $-$0.695/1.000. The disk shown has a radius of 12.5 $\ell$.}
\end{figure}

\begin{figure}
\centering
\includegraphics[width=0.45\textwidth,height=0.23\textwidth]{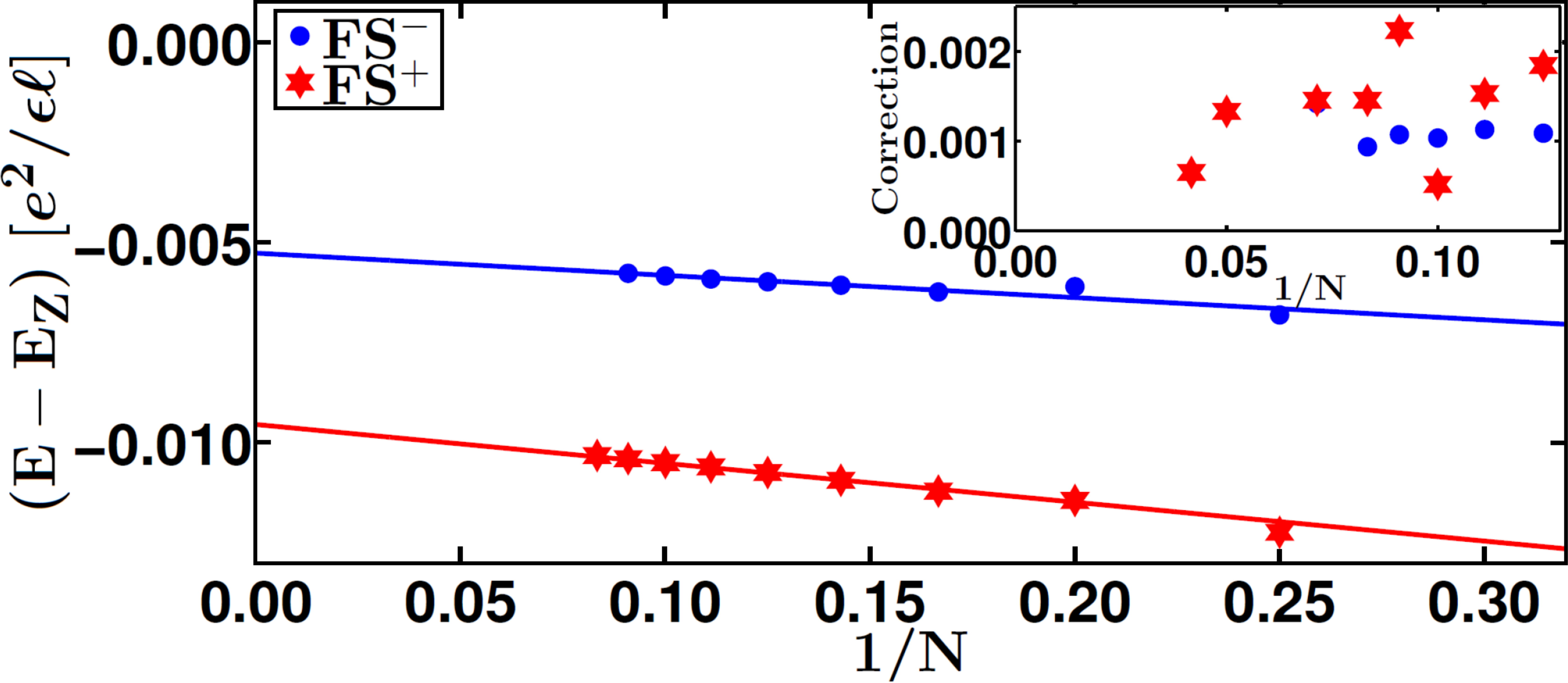}
\caption{\label{Fig4}{\bf Thermodynamic extrapolation of the binding energies of the fractional skyrmions.} The blue (red) symbols show the energies of negative (positive) fractional skyrmions for a system of $N$ particles with zero transverse width, obtained from exact diagonalization. The inset shows the amount by which finite width corrections lower the energy of the FS for a sample of width 33 nm. }
\end{figure}

\begin{figure}
\centering
\includegraphics[width=0.45\textwidth,height=0.23\textwidth]{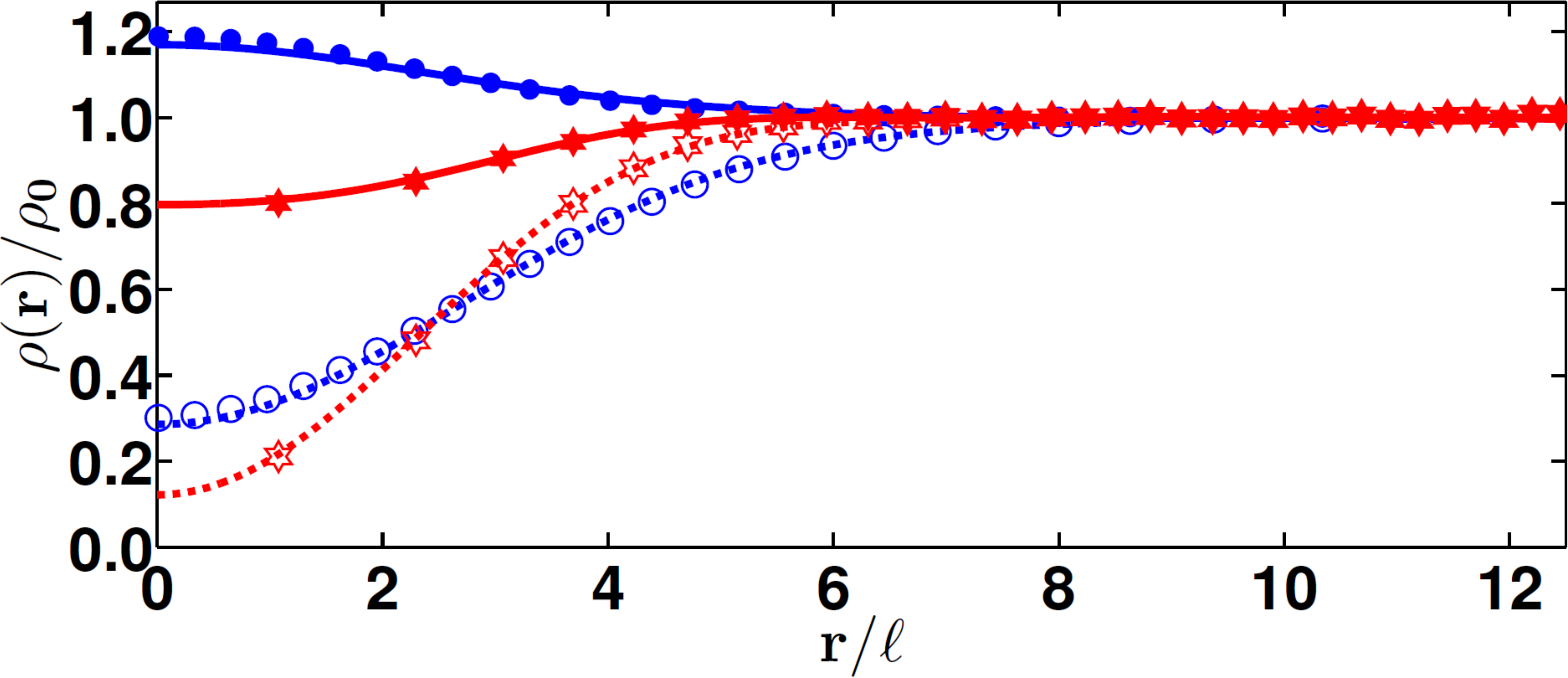}
\caption{\label{Fig5}{\bf Comparison of exact and CF density profiles for fractional skyrmions.} This figure shows the total density ($\rho$) and the density of spin-up particles ($\rho^\uparrow$) for FS$^{-}$ (blue) and FS$^{+}$ (red) obtained from exact (dotted and dashed lines) and CF diagonalization (filled and empty symbols). A near perfect overlay of the CF and exact curves shows that the wave function of the FS$^\pm$ obtained from CF diagonalization is almost identical to the exact one. The results are for 12 particles, and the density is quoted in units of the density of the uniform 1/3 state, denoted by $\rho_0$.}
\end{figure}

\begin{figure}
\centering
\includegraphics[width=0.5\textwidth]{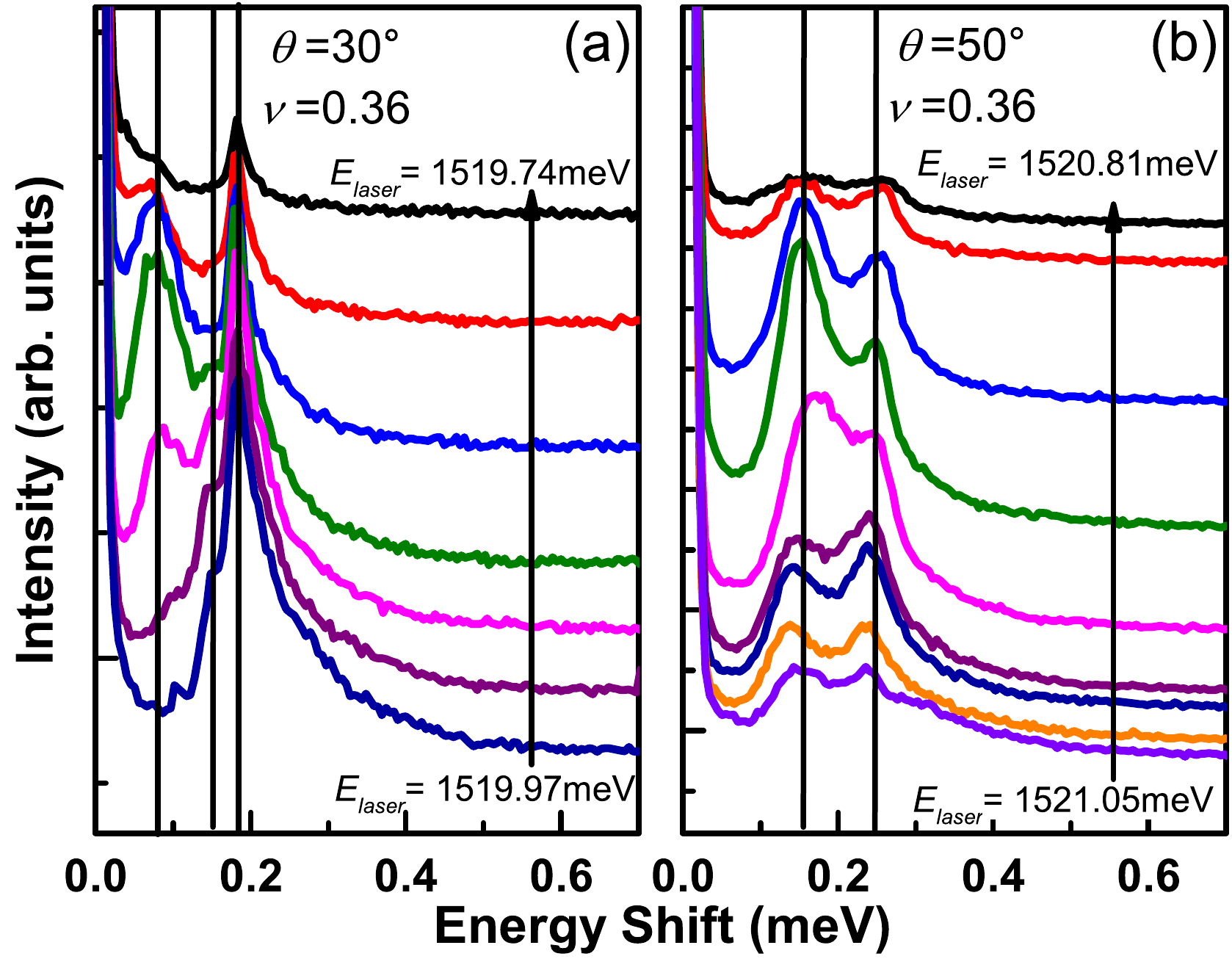}
\caption{\label{Fig6}{\bf Excitations in resonant inelastic light scattering spectra.} Panels (a) and (b) show typical spectra at $\nu=0.36$ for $\theta$=30$^{o}$ and $\theta$=50$^{o}$, respectively, as a function of the incident laser energy $E_{\rm laser}$.}
\end{figure}

\begin{figure}
\centering
\includegraphics[width=0.5\textwidth]{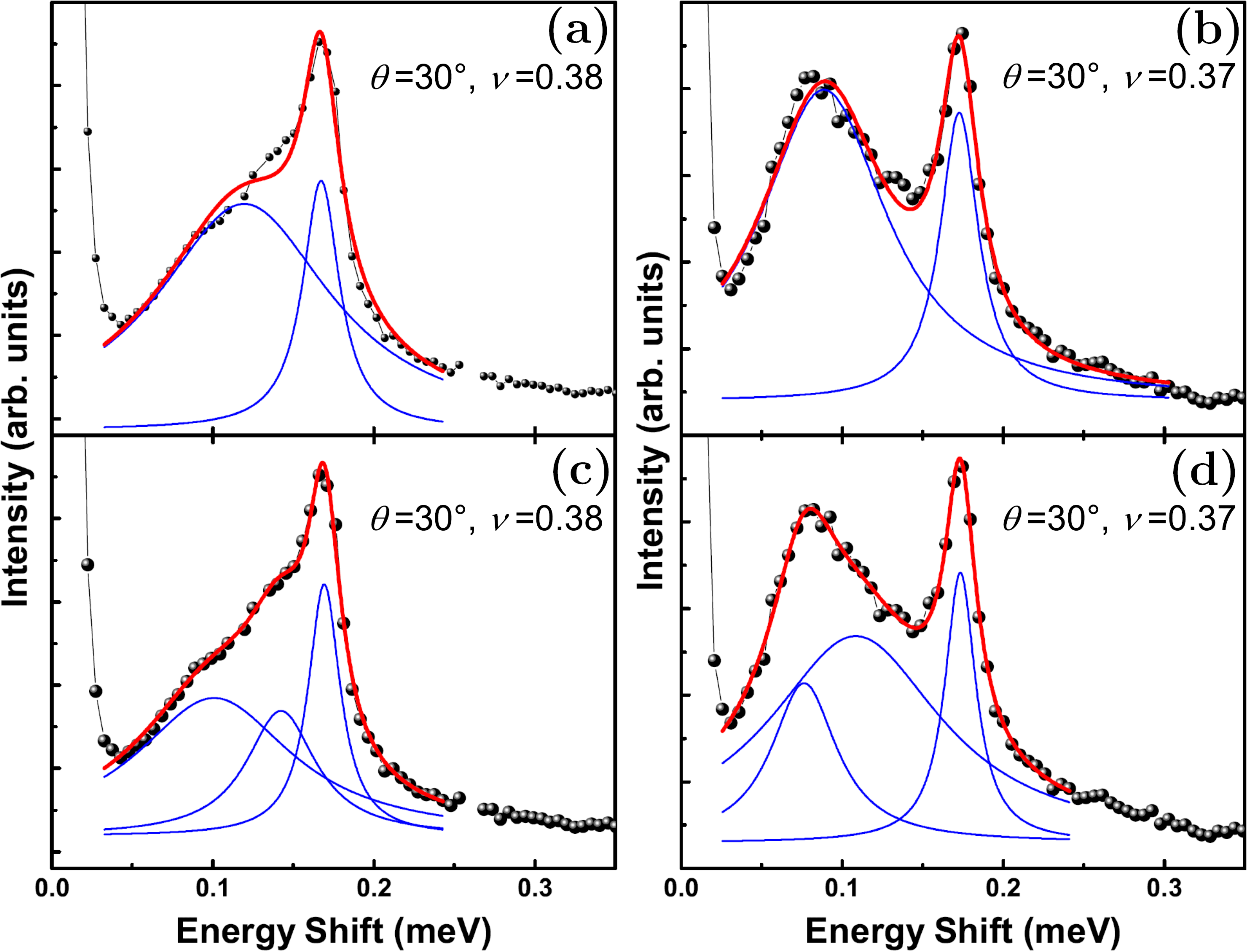}
\caption{\label{Fig7}{\bf Lorentzian fits to the RILS spectra.} RILS spectra obtained at $30^{\circ}$ tilt for $\nu=0.38$ (panels (a) and (c)) and $\nu=0.37$ (panels (b) and (d)). The raw RILS data are displayed as black bullets and the Lorentzian fits to the data as red solid line. The blue lines in each panel show the individual Lorentzians used to obtain the fit to the data. At $30^{\circ}$ tilt the data do not fit well to two Lorentzians as seen in panels (a) and (b), but fit well to three Lorentzians as shown in panels (c) and (d).} 
\end{figure}

We identify the mode just below the Zeeman energy with the minimal FS. This identification is supported by several observations.  At exactly $\nu=1/3$, no modes are observed below $E_{\rm Z}$, as expected. For $\nu\lesssim 1/3$, a single sub-$E_{\rm Z}$ mode is observed\cite{Gallais06}. The excellent quantitative agreement between theory and experiment seen in Fig.~\ref{Fig1}(j) confirms its identification with FS$^+$. We next consider $\nu\gtrsim1/3$. At 30$^\circ$ tilt, the energy of the mode slightly below $E_{\rm Z}$ is in excellent agreement with the calculated energy of the FS$^-$ (including finite thickness correction). We attribute the absence of this mode at 50$^\circ$ tilt to transition from a partially polarized ground state into a fully spin polarized ground state as $\kappa$ is raised from 0.018 to 0.023 by increasing the tilt. This is consistent with calculations\cite{Archer13b} that have shown that the ground state at $\nu\gtrsim 1/3$ has a transition from partially spin polarized state to a fully spin polarized state at $\kappa\approx 0.020$. We note that unlike their integral counterparts, the FS$^+$ and FS$^-$ are not related by particle-hole symmetry, as indicated by their different binding energies.

A discussion of various approximations is in order. LL mixing and disorder, not included in our calculations, are likely to provide small corrections. Calculations\cite{Scarola00} have shown that LL mixing is a minor effect, especially because finite width weakens the short range part of the interaction that is primarily responsible for causing admixture with higher LLs. Disorder is also known significantly to diminish the energy of a charged excitation. However, we are concerned here with the change in the energy of a charge-neutral SFE due to its binding with an already-present CF particle or hole, which we expect to be less sensitive to disorder\cite{Scarola00}. We have also neglected interaction between the SF and other CF particles or holes, which is valid only close to $\nu=1/3$ where the density of CF particles or holes is very small. The $\nu$ dependence of the energy of FS$^\pm$ indicates its renormalization due to the presence of other CF particles or holes in its vicinity.  The much weaker dependence of the measured dispersion of FS$^-$ indicates a weaker interaction between CF particles. The $\nu$ dependence of the FS energy can in principle allow a further investigation into the inter-CF interactions, although we have not pursued that here.

The excitation spectrum also contains skyrmions binding $K>2$ SFEs, with energy $KE_{\rm Z}-E^K_{\rm b}$. These lie in the continuum above $E_{\rm Z}$ for experimental parameters considered here. We expect that these couple weakly to RILS because they require the incident photon to excite $K\geq 2$ SFEs, a higher order scattering process.

\begin{figure}
\centering
\includegraphics[width=0.8\textwidth]{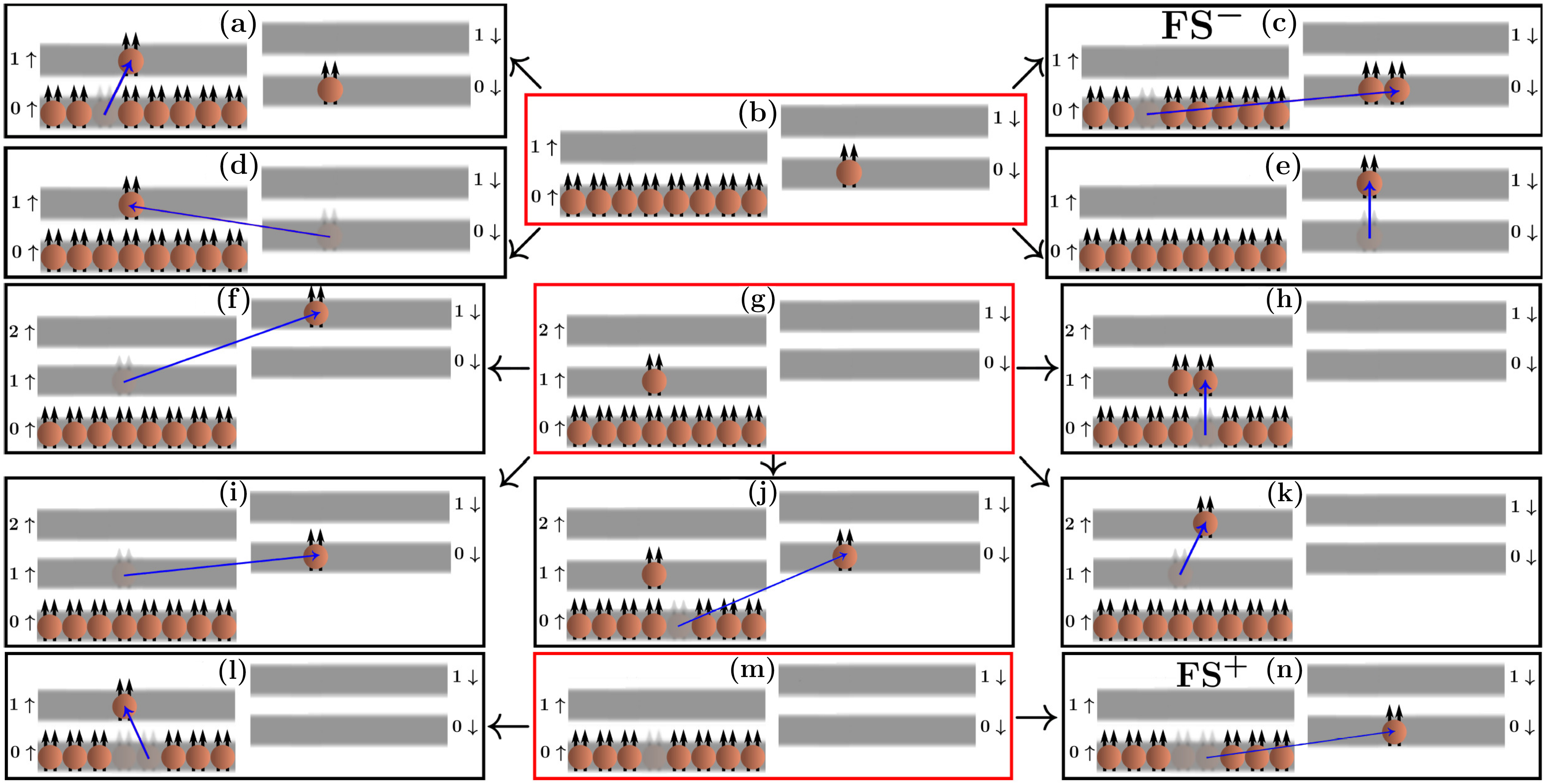}
\caption{\label{Fig8}{\bf Elementary excitations in the vicinity of filling factor $\nu=1/3$.} The panels (b) and (g) indicate the partially spin polarized and the fully spin polarized ground states  at $\nu \gtrsim 1/3$, and the panel (m) indicates the fully spin polarized ground state for $\nu\lesssim 1/3$. The other panels show the excitations obtained by either promoting (or demoting) a single composite fermion to a higher (lower) $\Lambda$ level or flipping its spin or doing both. The Panel (c) shows the FS$^{-}$ excitation and the Panel (n) the FS$^{+}$ excitation. }
\end{figure}

For completeness, we have considered excitations other than the FSs. Fig.~\ref{Fig8} shows schematically various elementary excitations of the states corresponding to panels (a), (c) and (e) of Fig.~\ref{Fig1}. We have performed an exhaustive study of all of these excitations by the standard methods of the CF theory \cite{Scarola00,Mandal01,Majumder09} as well as from an extrapolation of the results available from exact diagonalization.  The Coulomb contributions to their energy are given in Table \ref{tab:Table1} for both zero width and the width of 33 nm (which is close to the quantum well width in the experiment). The mode shown in Fig.~\ref{Fig8}(a) can be viewed as the mode in Fig.~\ref{Fig8}(d) plus a spin wave; given that the Coulomb energy of the spin wave goes to zero at small wave vector according to Larmor's theorem, these two have the same Coulomb energies (although they have different Zeeman contributions). For the modes shown in Fig.~\ref{Fig8}(c), Fig.~\ref{Fig8}(f), Fig.~\ref{Fig8}(i) and Fig.~\ref{Fig8}(j) (Fig.~\ref{Fig8}(d)) we need to add (subtract) the $E_{\rm Z}$ before comparing to the experiments. 

For $\nu<1/3$ only a single sub-$E_{\rm Z}$ mode (the FS$^+$) is expected from theory and only one such mode is observed experimentally. For $\nu>1/3$ other sub-$E_{\rm Z}$ modes are possible. In this region, using the parameters of Ref.~\cite{Dujovne03}, in addition to the FS$^{-}$, the mode shown in Fig.~\ref{Fig8}(d) lies below $E_{\rm Z}$ for the partially spin polarized state, and the mode shown in Fig.~\ref{Fig8}(i) lies below the $E_{\rm Z}$ for the fully spin polarized state. We assign the mode in panel (l) of Fig.~\ref{Fig1} to the excitation shown in Fig.~\ref{Fig8}(i). The Coulomb energy contribution to the measured energy, $\sim -$0.013 $e^2/\epsilon \ell$, is to be compared with the theoretical Coulomb energy (including finite width correction) of $-$0.018 $e^2/\epsilon \ell$. We further assign the mode in panel (k) of Fig.~\ref{Fig1} indicated by green stars to the excitation shown in Fig.~\ref{Fig8}(d). The Coulomb contribution to the energy of this mode is theoretically calculated to be 0.018 $e^2/\epsilon \ell$ whereas the measured one is 0.030 $e^2/\epsilon \ell$. We find this level of agreement acceptable, considering that both these numbers are differences between the energies of two single particle excitations, which are known to be sensitive to the effects of disorder.

RILS can also detect magneto-roton modes, which are particle hole excitations of composite fermions (to be distinguished from the minimal FSs that are bound state of three composite fermions). The magneto-roton modes of the 1/3 or 2/5 states occur at energies much higher than $E_{\rm Z}$ for the parameters of the experiments in question. Additional low energy magneto-rotons are in principle possible at incompressible FQHE states of composite fermions in the range $1/3<\nu<2/5$, such as 4/11, 5/13 and 3/8 \cite{Pan03,Liu14a,Balram15}, which correspond to neutral spin-conserving modes of the composite fermions in the minority spin sector\cite{Chang03b,Wojs04,Mukherjee15}.  Several observations indicate that this physics is not relevant to the mode identified above as FS$^-$. First, the CF-FQHE states at 4/11, 5/13 and 3/8 are stabilized at temperatures below the minimum temperature (50mK) of the experiments of Fig.~\ref{Fig1}. Second, the experimental mode identified as FS$^{-}$ does not require a fine tuning of $\nu$. Third, the energy of the observed mode agrees with the theoretically calculated energy of FS$^-$. Finally, the neutral magnetoroton modes of 4/11, 3/8 and 5/13 are expected to occur at $\sim$0.002 $e^2/\epsilon \ell$  \cite{Mukherjee14,Mukherjee12}, which is much lower than the energy of FS$^-$.

In conclusion, we have shown that the smallest version of the fractionally charged skyrmions can be created optically in the vicinity of $\nu=1/3$ when a photoexcited neutral exciton forms a bound state with an already present charged CF particle or hole. Furthermore, we provide strong evidence that the modes observed slightly below the long wave length spin wave mode at the Zeeman energy are precisely these skyrmions. In particular, the measured binding energy of these skyrmions is in excellent agreement with theory. To our knowledge, this is to date the best agreement between theory and experiment for a non-trivial excitation in the FQHE. The study of skyrmions for $\nu\approx 1/3$ provides a sensitive probe into the inter-composite fermion interaction, and also sheds light on the spin polarization of the ground state.

\begin{table}
\centering
\begin{tabular}{|c|c|c|c|}
\hline
\multicolumn{1}{|c|}{Mode} & \multicolumn{2}{|c|}{width, $w=0$} & \multicolumn{1}{|c|}{$E_{\text{$w=33$ nm}}-E_{w=0}$} \\ \hline
	        & exact		  	& 	CF theory			&	CF theory	 \\ \hline
(a)+(e),(d)	& 0.0284(3)		&	0.0258(0)			& -0.0073(0)  \\ \hline
(c)  		& -0.0052(2)		&	$\sim$ -0.0059(7)		& 0.0010(0)  \\ \hline
(f)+(j)      	& 0			&	0				& 0	     \\ \hline
(h)      	& 0.0422(33)		&	$\sim$ 0.0436(113)		& -0.0095(207) \\ \hline
(i)      	& -0.0284(3)		&	-0.0258(0)			& 0.0073(0)  \\ \hline
(k)      	& -			&	0.0867(1)			& -0.0224(1)  \\ \hline
(l)     	& 0.0369(17)		&	0.0366(47)			& -0.0117(77) \\ \hline
(n)     	& -0.0096(2)		&	$\sim$ -0.0108(24)		& 0.0013(7)  \\ \hline
\end{tabular}
\caption {{\bf Energy of the elementary excitations in the vicinity of $\nu=1/3$ shown in Fig. \ref{Fig8}.} The Coulomb energy of the the elementary excitations near $\nu=1/3$ determined by an extrapolation of the finite system results, obtained by exact diagonalization (second column) and the CF theory (third column), for quantum well width $w=0$. The last column gives the difference in the energies of each mode for quantum wells of widths $w=33$ nm and $w=0$, obtained by the CF theory. All energies are quoted in units of $e^2/\epsilon \ell$. The cases where linear extrapolation in $1/N$ to the thermodynamic limit is not very accurate are marked by the symbol $\sim$ to indicate larger uncertainty. The total energy of (c), (i) and (n) ((d)) is obtained by adding (subtracting) the Zeeman splitting $E_{\rm Z}$ as explained in the text. (Note: A combination of (a) and (e) is needed to obtain $S^2$ eigenstates; the same is true of (f) and (j). (k) is an eigenstate of the Hamiltonian but it is in general an excited state.)}
\label{tab:Table1} 
\end{table}

{\em Acknowledgements}: We acknowledge financial support from the U.S. Department of Energy, Office of Science, Basic Energy Sciences, under Award No. DE-SC0005042 (ACB,JKJ), from the US National Science Foundation under grant DMR-1306976 (AP), from the National Science Centre, Poland, under grant 2014/14/A/ST3/00654 (AW). We thank Research Computing and Cyberinfrastructure at Pennsylvania State University (supported in part through instrumentation funded by the National Science Foundation through grant OCI-0821527), Wroc{\l}aw Centre for Networking and Supercomputing and Academic Computer Centre CYFRONET, both parts of PL-Grid Infrastructure, and the DFG excellence program Nanosystems Initiative Munich (UW). We thank Indian Institute of Science Education and Research, Pune for their kind hospitality during the final stages of this work.\\
{\em Author contributions}: A.P. and J.K.J. planned and organized this project; U.W. and A.P. analyzed the experimental data; A.C.B. and A.W performed theoretical calculations; all authors discussed the results and contributed to the writing of the manuscript. \\
{\em Competing financial interests}: The authors declare no competing financial interests. \\
{\em Correspondence}: Correspondence and requests for materials should be addressed to J.K.J.~(email: jkj2@psu.edu).\\

\bibliography{../../Latex-Revtex-etc./biblio_fqhe}
\bibliographystyle{apsrev}

\end{document}